\documentstyle[preprint,aps,floats,epsf]{revtex}
\begin{document}
\draft
\title{Is the Polarized Antiquark Sea in the Nucleon Flavor
Symmetric?}
\author{R.S.~Bhalerao}
\address{Department of Theoretical Physics, 
Tata Institute of Fundamental Research,\\
Homi Bhabha Road, Colaba, Mumbai 400 005, India}

\maketitle

\begin{abstract}
We show that the model which naturally explains the $\bar u \ne
\bar d$ asymmetry in the nucleon and is in quantitative agreement
with the Gottfried sum rule data, also predicts that in the
proton $\Delta \bar u > 0 > \Delta \bar s > \Delta \bar d$ and
$\Delta \bar u - \Delta \bar d > \bar d - \bar u > 0$. At the
input scale, these results can be derived even analytically. Thus
the violation of the flavor symmetry is more serious in the
polarized case than in the unpolarized case. In contrast, many
recent analyses of the polarized data have made a simplifying
assumption that all the three $\Delta \bar q$'s have the same
sign and magnitude. We point out the need to redo these analyses,
allowing for the alternate scenario as described above. We
present predictions of the model for the $W^-$ asymmetry in
polarized $pp$ scattering, which can be tested at RHIC;
these are quite different from those available in the literature.
\end{abstract}




\newpage

Several comprehensive analyses of the polarized deep inelastic
scattering (DIS) data, based on next-to-leading-order quantum
chromodynamics (QCD), have appeared recently
\cite{str99,glu96,geh96,abe97,lea99,bou98,flo98,got00,gor98,gho00,alt97}.
In these analyses the polarized parton
density functions (PDFs) are either written in terms of the
well-known parameterizations of the {\it unpolarized} PDFs or
parameterized independently, and the unknown parameters are
determined by fitting the polarized DIS data. Additional
simplifying assumptions are often made; the one that has been
widely used in the literature \cite
{str99,glu96,geh96,abe97,lea99,bou98,flo98,got00} is
\begin{equation}
\Delta \bar u = \Delta \bar d = \lambda \Delta \bar s,
\end{equation}
with a positive $\lambda$ which is usually set equal to unity.
Recently the HERMES and SMC collaborations \cite{ack99,ade98b}
too analyzed their
inclusive and semi-inclusive DIS data assuming all $\Delta \bar
q$'s to be of the same sign. The same assumption has also been
used to make predictions for future accelerators; see
e.g. \cite{kum99}. In some analyses \cite{alt97}, the nonsinglet PDFs
$\Delta q_3$ and $\Delta q_8$ are assumed to differ only by a
constant multiplicative factor (see (21) below).

In this paper, we examine these simplifying assumptions made in
the literature. This is important because a similar ad hoc
assumption about the flavor decomposition of the unpolarized
antiquark sea, $\bar u = \bar d$, turned out to be wrong when
accurate data on muon DIS became available \cite{arn94}, and the
global analyses of the unpolarized data had to be redone. Here we
derive a series of inequalities satisfied by the PDFs and point
out the need to redo the global analyses of the polarized DIS
data in the light of these inequalities, allowing in particular
for the violation of the flavor symmetry in the polarized
antiquark sea; see (20) below. We then present predictions of our
model, which can be tested in polarized $pp$ scattering at RHIC,
BNL. Finally, we describe other recent works on flavor asymmetry of
polarized sea distributions.

We use the framework of the statistical model for polarized and
unpolarized structure functions and PDFs of the proton and the
neutron, which was presented recently \cite{bha96,bha00}. This
model provided a natural explanation of the $\bar u \ne \bar d$
asymmetry in the nucleon and was in quantitative agreement with
the Gottfried sum rule data. Additionally, it reproduced the data
on $F_2^p(x,Q^2)$ for $0.00001<x<1$ and $2.5<Q^2<5000$ GeV$^2$,
$F_2^p(x) - F_2^n(x),~ F_2^n(x) / F_2^p(x),~ xg(x),~ {\bar d}(x)
- {\bar u}(x),~ d(x)/u(x)$, the fractional momentum of charged
partons and the polarized structure functions $g_1^{p,n}(x)$, at
various $Q^2$. Out of those, only the $F_2^p$ and $(F_2^p -
F_2^n)$ data, both at $Q^2 = 4$ GeV$^2$ only,
were used as an input to fix
the model parameters, and all other results served as model
predictions. In particular, the $d(x)/u(x)$ ratio in the limit $x
\rightarrow 1$ turned out to be 0.22 in good agreement with the
QCD prediction 0.2 \cite{far75}. At the input scale
($Q^2=Q_0^2=M^2$,
where $M$ is the nucleon mass), 
all $xq(x)$ and $x \bar q(x)$ distributions were
found to be valence-like, and $xg(x)$ was found to be constant in
the limit $x \rightarrow 0$. {\it Thus the total number of gluons
was logarithmically divergent providing a strong a posteriori
justification for the statistical model ansatz} \cite{bha00}.
Contrary to the common practice, the polarized and unpolarized
data were reproduced in a single framework and the simplifying
assumption of charge symmetry was not made. Here we further
explore the predictive power of the model.

If $n_{\alpha(\bar \alpha)\uparrow(\downarrow)}$ denotes the
number of quarks (antiquarks) of flavor $\alpha$ and spin
parallel (antiparallel) to the proton spin, then any model of
PDFs in the proton has to satisfy the following constraints:
\begin{eqnarray}
n_{u\uparrow}+n_{u\downarrow}-n_{\bar u \uparrow}-n_{\bar u
\downarrow}
&=&2,\\
n_{d\uparrow}+n_{d\downarrow}-n_{\bar d \uparrow}-n_{\bar d
\downarrow}
&=&1,\\
n_{s\uparrow}+n_{s\downarrow}-n_{\bar s \uparrow}-n_{\bar s
\downarrow}
&=&0,\\
n_{u\uparrow}-n_{u\downarrow}+n_{\bar u \uparrow}-n_{\bar u
\downarrow}
&=& \Delta u + \Delta \bar u,\\
n_{d\uparrow}-n_{d\downarrow}+n_{\bar d \uparrow}-n_{\bar d
\downarrow}
&=& \Delta d + \Delta \bar d,\\
n_{s\uparrow}-n_{s\downarrow}+n_{\bar s \uparrow}-n_{\bar s
\downarrow}
&=& \Delta s + \Delta \bar s.
\end{eqnarray}
The RHSs of (5)-(7) have been measured by several groups. We use
$(\Delta u + \Delta \bar u) =  0.83 \pm 0.03, 
~(\Delta d + \Delta \bar d) = -0.43 \pm 0.03, 
~(\Delta s + \Delta \bar s) = -0.10 \pm 0.03$;
see \cite{abe98}. The parton numbers $n_{\alpha(\bar
\alpha)\uparrow(\downarrow)}$ in (2)-(7) are obtained by
integrating the appropriate number density $dn/dx$ over $x$. The
various $\Delta$'s are also $x$-integrated quantities.

The RHSs of (2)-(4) are clearly $Q^2$-independent. The RHSs of
(5)-(7) are also $Q^2$-independent in the jet and Adler-Bardeen
(AB) schemes: Recall that the nonsinglets $\Delta q_3 = (\Delta u
+ \Delta \bar u) - (\Delta d + \Delta \bar d)$ and $\Delta q_8 =
(\Delta u + \Delta \bar u) + (\Delta d + \Delta \bar d) -2
(\Delta s + \Delta \bar s)$ are $Q^2$-independent in all
renormalization schemes because of the conservation of the
nonsinglet axial vector current, and the singlet $\Delta \Sigma =
(\Delta u + \Delta \bar u) + (\Delta d + \Delta \bar d) + (\Delta
s + \Delta \bar s)$ is $Q^2$-independent in the jet and AB
schemes because of the Adler-Bardeen theorem \cite{adl69}. As a
result, $(\Delta u + \Delta \bar u), ~(\Delta d + \Delta \bar d)$
and $(\Delta s + \Delta \bar s)$ which can be expressed as linear
combinations of $\Delta q_3, ~\Delta q_8$ and $\Delta \Sigma$,
are also $Q^2$-independent in these two schemes. In the
$\overline {MS}$ scheme, on the other hand, $\Delta \Sigma$ is
$Q^2$-independent at the leading order and only weakly
$Q^2$-dependent at the next-to-leading order. Empirically
too $\Delta \Sigma$ is found to be almost $Q^2$-independent; see
e.g. Fig. 5 of \cite{str97}. Hence in the $\overline {MS}$ scheme the
RHSs of (5)-(7) are expected to be nearly $Q^2$-independent.

We now show how the statistical model naturally leads to a
violation of the flavor symmetry in the unpolarized and polarized
seas in the nucleon. Consider the following 6 equations:
\begin{eqnarray}
2 n_{u\uparrow}  -2 n_{\bar u \downarrow} &=& 2.83, \eqnum{8}\\
2 n_{u\downarrow}-2 n_{\bar u \uparrow}   &=& 1.17, \eqnum{9}\\
2 n_{d\uparrow}  -2 n_{\bar d \downarrow} &=& 0.57, \eqnum{10}\\
2 n_{d\downarrow}-2 n_{\bar d \uparrow}   &=& 1.43, \eqnum{11}\\
2 n_{s\uparrow}  -2 n_{\bar s \downarrow} &=&-0.10, \eqnum{12}\\
2 n_{s\downarrow}-2 n_{\bar s \uparrow}   &=& 0.10. \eqnum{13}
\end{eqnarray}
These are obtained from (2)-(7) by linearly combining the latter
set of equations in pairs. E.g. (8) and (9) are obtained by
adding or subtracting (2) and (5).

It was shown in \cite{bha96} that the parton number density
$dn/dx$ in the infinite-momentum frame, at the input scale, is
given by
\begin{equation}
{dn \over dx} = {M^2 x \over 2} \int ^{M/2}_{xM/2} {dE \over
E^2}~~ {dn \over dE}, \eqnum{14a}
\end{equation}
where
\begin{equation} 
dn/dE = g ~f(E)~ (VE^2/2\pi^2 + aR^2E + bR),
\eqnum{14b}
\end{equation}
is the density in the nucleon rest frame.  Here $M$ is the
nucleon mass, $E$ is the parton energy in the nucleon rest frame,
$g$ is the spin-color degeneracy factor, $f(E)$ is the usual
Fermi or Bose distribution function $f(E) = \{\exp[(E-\mu)/T]\pm
1\}^{-1}$, $V$ is the nucleon volume and $R$ is the radius of a
sphere with volume $V$. The three terms in (14b) are the volume,
surface and curvature terms, respectively; in the thermodynamic
limit only the first survives. The two free parameters $a$ and
$b$ in (14b) were determined in \cite{bha00} by fitting the
structure function $F_2(x,Q^2)$ data at $Q^2 = 4$ GeV$^2$. Their
values as well as the values of the temperature $(T)$ and
chemical potential $(\mu)$ which get determined due to (2)-(7),
were given in \cite{bha00}.

At the input scale, with the help of (14), (8) can be written in
a full form as
\begin{equation} 
\int^1_0 dx {M^2 x \over 2} \int ^{M/2}_{xM/2} {dE \over E^2}~~ 
g~ (VE^2/2\pi^2 + aR^2E + bR) 
\left[ {2 \over e^{\beta(E-\mu_{u \uparrow})} + 1}
 -     {2 \over e^{\beta(E-\mu_{\bar u \downarrow})} + 1} \right]
=2.83 ~.
\eqnum{15} 
\end{equation}
It is straightforward to show that the chemical potentials for
quarks and antiquarks satisfy the relations
\begin{eqnarray}
\mu_{\bar q \uparrow}   &=& - \mu_{q \downarrow}, \eqnum{16a}\\
\mu_{\bar q \downarrow} &=& - \mu_{q \uparrow}.   \eqnum{16b}
\end{eqnarray}
So it follows from (15) and (16b) that $\mu_{u \uparrow} >
0$. Similar arguments show that $\mu_{u \downarrow}, ~\mu_{d
\uparrow}, ~\mu_{d \downarrow}$ and $\mu_{s \downarrow}$ are
positive and $\mu_{s \uparrow}$ is negative. Moreover, since the
RHSs of (12) and (13) differ only in sign, we have $\mu_{s
\uparrow}= -\mu_{s \downarrow}$. Since RHSs of (8)-(13) can be
arranged as $2.83 > 1.43 > 1.17 > 0.57 > 0.10 > -0.10$, the
corresponding chemical potentials satisfy
\begin{equation}
\mu_{u \uparrow} > \mu_{d \downarrow} > \mu_{u \downarrow} >
\mu_{d \uparrow} > (\mu_{s \downarrow} = \mu_{\bar s \downarrow})
> 0 > (\mu_{s \uparrow} = \mu_{\bar s \uparrow}) > \mu_{\bar d
\downarrow} > \mu_{\bar u \uparrow} > \mu_{\bar d \uparrow} >
\mu_{\bar u \downarrow}.  \eqnum{17}
\end{equation}
It will be useful to recall the actual values of the $\mu$'s
given in \cite{bha00}. They are (in MeV) $ \mu_{u \uparrow} =
210, ~\mu_{d \downarrow} = 106, ~\mu_{u \downarrow} = 86, ~\mu_{d
\uparrow} = 42, ~\mu_{s \downarrow} = 7, ~\mu_{s \uparrow} = -7.$
$\mu$'s for the antiquarks follow from (16). [The RHSs of
(8)-(13) are sufficiently different from each other so that the
experimental errors in $(\Delta q + \Delta \bar q)$, quoted
above, will not alter the ordering in (17).] (17) together with
(14) yields, {\it at the input scale} $Q_0^2~~(=M^2= 0.88$ GeV$^2$):
\begin{equation}
n_{u \uparrow} > n_{d \downarrow} > n_{u \downarrow} > n_{d
\uparrow} > (n_{s \downarrow} = n_{\bar s \downarrow}) > (n_{s
\uparrow} = n_{\bar s \uparrow}) > n_{\bar d \downarrow} >
n_{\bar u \uparrow} > n_{\bar d \uparrow} > n_{\bar u \downarrow}
> 0.  \eqnum{18}
\end{equation}
As a check, it is easy to verify that (18) reproduces the correct
signs of the RHSs of (2)-(7). Notice the symmetric arrangement of
the $\mu$'s in (17) and the consequent arrangement of the $n$'s
in (18).

{\it To recapitulate, the statistical model provides a
quantitative method to incorporate the effects of the Pauli
exclusion principle into the PDFs: the RHSs of the number
constraints} (2)-(7) {\it or equivalently} (8)-(13), {\it force
the various chemical potentials and hence the parton
distributions to be arranged as in} (17) {\it and} (18), {\it
respectively, at the input scale}. 

Further consequences of (18) are easy to derive: (Note $n_q =
n_{q \uparrow} + n_{q \downarrow}$ and $\Delta q = n_{q \uparrow}
- n_{q \downarrow}$.)

\noindent (a) The general positivity constraints on the polarized
and unpolarized PDFs: $|\Delta q| \leq n_q$ are satisfied
trivially.

\noindent (b) $\Delta u > 0, ~\Delta d < 0, ~\Delta s < 0$.

\noindent (c) $\Delta \bar u > 0, ~\Delta \bar d < 0, ~\Delta
\bar s < 0$. This is in contrast to the assumption (1) made in
the literature
\cite{str99,glu96,geh96,abe97,lea99,bou98,flo98,got00,ack99,ade98b,kum99}
that all the three $\Delta
\bar q$'s have the same sign.

\noindent (d) $\Delta u_v = \Delta u - \Delta \bar u = n_{u
\uparrow} - n_{u \downarrow} - n_{\bar u \uparrow} + n_{\bar u
\downarrow} > 0$, because the two $n_{\bar u}$ terms are too
small compared to the two $n_u$ terms (see (18)) to change the
sign of the RHS.

\noindent (e) $\Delta d_v = \Delta d - \Delta \bar d = n_{d
\uparrow} - n_{d \downarrow} - n_{\bar d \uparrow} + n_{\bar d
\downarrow} < 0$, because the two $n_{\bar d}$ terms are too
small compared to the two $n_d$ terms (see (18)) to change the
sign of the RHS.

\noindent (f) $\Delta s_v = \Delta s - \Delta \bar s = 0$.

\noindent (g) $\Delta q_3 = (\Delta u + \Delta \bar u) - (\Delta
d + \Delta \bar d) > 0$; see (b)-(c).

\noindent (h) $n_{\bar d} > n_{\bar u}$ which leads to the
Gottfried sum rule violation. {\it Thus the statistical model
naturally leads to the $\bar u \ne \bar d$ asymmetry in the
unpolarized sea} \cite{bha96}. Moreover, it was shown in \cite{bha00} 
that the
model is in {\it quantitative} agreement with the data on $(F_2^p
- F_2^n)$ vs $x$ and the Gottfried sum $S_G$.

\noindent (i) $\Delta \bar u - \Delta \bar d > n_{\bar d} -
n_{\bar u} > 0$. {\it Thus the violation of the flavor symmetry
is more serious in the polarized case than in the unpolarized
case.}

\noindent (j) $\Delta d - \Delta s = n_{d \uparrow} - n_{d
\downarrow} - n_{s \uparrow} + n_{s \downarrow} < 0$, because
$n_{s \uparrow}$ and $n_{s \downarrow}$ tend to cancel each
other, unlike $n_{d \uparrow}$ and $n_{d \downarrow}$. Combining
this result with (b) above, one gets $|\Delta d| > |\Delta s|$,
and
\begin{equation}
\Delta u > 0 > \Delta s > \Delta d.
\eqnum{19}
\end{equation}

\noindent (k) $\Delta \bar d - \Delta \bar s = n_{\bar d
\uparrow} - n_{\bar d \downarrow} - n_{\bar s \uparrow} + n_{\bar
s \downarrow} < 0$, because $n_{\bar s \uparrow}$ and $n_{\bar s
\downarrow}$ tend to cancel each other, unlike $n_{\bar d
\uparrow}$ and $n_{\bar d \downarrow}$. Combining this result
with (c) above, one gets $|\Delta \bar d| > |\Delta \bar s|$, and
\begin{equation}
\Delta \bar u > 0 > \Delta \bar s > \Delta \bar d.
\eqnum{20}
\end{equation}

We have derived the results (a)-(k) analytically, at the input
scale. They are borne out by actual numerical calculations; see
Fig. 1 which shows our polarized PDFs at the input scale
$Q_0^2=M^2=0.88$ GeV$^2$. We have evolved our polarized PDFs in the
next-to-leading-order QCD, in the $\overline {MS}$ scheme, in the
range $Q_0^2 < Q^2 < 6500$ GeV$^2$. We find that the results
(a)-(k) are valid throughout this range. 
Figure 2 shows that the violation of the flavor
symmetry is more serious in the polarized case than in the
unpolarized case, throughout this range.

Incidentally, we have examined another simplifying assumption
made e.g. in \cite{alt97}, namely
\begin{equation}
\Delta q_3(x,Q^2) = C~\Delta q_8(x,Q^2),
\eqnum{21}
\end{equation}
where C is a constant independent of $x$ and $Q^2$. The present
model predicts that (21) is not justified (Fig. 1).

The statistical model makes concrete predictions for various
asymmetries in polarized $pp$ scattering, which can be tested at
RHIC. For example, parity-violating single- and double-spin
asymmetries for $W$ production in the reactions $\overrightarrow{p}p
\rightarrow W^{\pm}X$ and $\overrightarrow{p} \overrightarrow{p}
\rightarrow W^{\pm}X$ respectively, are given by \cite{bou93,sof98}
\begin{eqnarray}
A_L^{PV}(W^+) &=& \frac {\Delta u(x_a,M_W^2) ~\bar d(x_b,M_W^2)-
\Delta \bar d(x_a,M_W^2) ~u(x_b,M_W^2)}
{u(x_a,M_W^2)~\bar d(x_b,M_W^2) + \bar d(x_a,M_W^2)~u(x_b,M_W^2)},
\eqnum{22}\\
A_L^{PV}(W^-) &=& \frac {-\Delta \bar u ~d + \Delta d~\bar u}
{\bar u ~d + d ~\bar u},
\eqnum{23}\\
A_{LL}^{PV}(W^+) &=& \frac {\Delta u ~\bar d -u ~\Delta \bar d
-\Delta \bar d ~u + \bar d ~\Delta u}{u~\bar d - \Delta u ~\Delta 
\bar d +\bar d ~u - \Delta \bar d ~\Delta u},
\eqnum{24}\\
A_{LL}^{PV}(W^-) &=& \frac {\bar u~\Delta d - \Delta \bar u~d - d
~\Delta \bar u + \Delta d ~\bar u} {\bar u ~d - \Delta \bar u 
~\Delta d + d~ \bar u - \Delta d~ \Delta \bar u},
\eqnum{25}
\end{eqnarray}
where $x_a= \sqrt {\tau} e^y, ~x_b= \sqrt {\tau} e^{-y}, ~\tau =
M_W^2/s$, ~$y$ is the rapidity of $W$ and $\sqrt{s}$ is the $pp$
center-of-mass energy. The arguments $x_a, ~x_b$ and $M_W^2$ are
suppressed in (23)-(25) for brevity of notation. 

In the present model, $\Delta u$ and $\Delta \bar u$ are positive and
$\Delta d$ and $\Delta \bar d$ are negative (see (b), (c) and
Fig. 1). Also note that $\Delta u \le u$ and $|\Delta \bar d| \le \bar
d$ (see (a) above). Hence it is straightforward to show that $0 <
A_L^{PV}(W^+) < 1$. Similarly, $-1 < A_L^{PV}(W^-) < 0,
~A_{LL}^{PV}(W^+)>0, ~A_{LL}^{PV}(W^-)<0.$ It is somewhat tedious but
again straightforward to show using (22)-(25) that $A_{LL}^{PV}(W^+) >
A_{L}^{PV}(W^+)$ and $|A_{LL}^{PV}(W^-)| > |A_{L}^{PV}(W^-)|$. 
A quick and crude way to convince oneself that $A_{LL}^{PV}(W^+) >
A_{L}^{PV}(W^+)$ is to ignore the (small) ``$\Delta \Delta$'' terms in
the denominator of (24), which makes the denominators of (22) and (24)
identical, and then to compare their numerators. In fact, at $y=0$ (or
$x_a=x_b$), ~$A_{LL}^{PV}(W^+)$ is seen to be almost twice as big as
$A_{L}^{PV}(W^+)$.

Figure 3 shows our predictions for $A_L^{PV}$ and $A_{LL}^{PV}$ for
$W^-$ production in polarized $pp$ scattering at $\sqrt s= 500$ GeV,
as a function of the rapidity $y$. The above inequalities for
$A_{L}^{PV}(W^-)$ and $A_{LL}^{PV}(W^-)$, which we derived
analytically here are borne out by the actual numerical results in
Fig. 3. Also shown for comparison are results reported in
\cite{sof98}. These are based on the parameterizations of polarized
PDFs given in \cite{glu96,geh96,bou95}. Asymmetries for $W^-$
production are sensitive to the sign of $\Delta \bar u$ which is
positive in the present model, negative in \cite{glu96,bou95} and
$x$-dependent in \cite{geh96}. The recent work of de Florian and
Sassot \cite{flo00} has yielded a clear preference for a positive
$\Delta \bar u$ distribution.

As stated earlier, the HERMES and SMC collaborations
\cite{ack99,ade98b} analyzed their inclusive and semi-inclusive DIS
data assuming all $\Delta \bar q$'s to be of the same sign. Recently,
Morii and Yamanishi \cite{mor00} have reanalyzed these data and have
estimated $\Delta \bar d(x) - \Delta \bar u(x)$ at $Q^2 = 4$
GeV$^2$. It is evident from their Fig. 1 that $\Delta \bar u(x) -
\Delta \bar d(x)$ is positive and has a peak at $x \simeq 0.06$ where
$x(\Delta \bar u(x) - \Delta \bar d(x))$ is $\simeq 0.05$. All these
observations are consistent with our Fig. 2.

Another model which is able to generate flavor asymmetric polarized
antiquark sea is the chiral quark soliton model (CQSM)
\cite{dia96,pob99,dre98,dre99a,dre99b,wak99}. Results in our Fig. 2
are strikingly similar to those in \cite{dia96,pob99,dre99a}.
This is remarkable because the physics
inputs of the two models are quite different. It is also noteworthy
that the origin of the $\bar u \ne \bar d$ and $\Delta \bar u \ne
\Delta \bar d$ asymmetries is quite simple in the statistical model.
While the role of gluons is yet to be understood in CQSM, the
statistical model predicts a positive $\Delta g(x,Q^2)$. The pion
cloud model also gives rise to the $\bar u \ne \bar d$ asymmetry, [for
a recent review, see \cite{kum98}], and there have been some attempts
to generate polarization by including spin-1 resonances in that
model. These attempts have been commented upon in
\cite{dre99a,dre99b}. Recently, Gl\"uck and Reya \cite{glu00} have
discussed the issue of flavor asymmetry, in a phenomenological way
making use of the Pauli exclusion principle. We recall that the
statistical model \cite{bha96,bha00} provides a quantitative method to
incorporate the effects of the Pauli exclusion principle into the
PDFs.

We have treated all partons as massless: $m_u = m_d = m_s =
0$. If $m_s$ is taken to be nonzero, then (14) will have to be
generalized, but the parton densities still have to satisfy
(2)-(7) and equivalently (8)-(13). So it is not obvious how that
will affect the symmetric arrangement of the $\mu$'s in (17) and
the consequent arrangement of the $n$'s in (18), at the input
scale. This is a  nontrivial problem which needs to be
investigated further.

In conclusion, we have derived, on rather general grounds, a
series of inequalities for the polarized PDFs; see (a)-(k) above.
This points to the need to redo the analyses
\cite{str99,glu96,geh96,abe97,lea99,bou98,flo98,got00,ack99,ade98b} of
polarized data, allowing for the alternate scenario as in
(19)-(20).  Some of the inequalities can be tested in the
forthcoming spin-physics program at RHIC, BNL.  To illustrate, we
have given our predictions for the $W^-$ asymmetries; these are quite
different from those available in the literature.

\bigskip\bigskip

I would like to acknowledge the hospitality of the Nuclear Theory
Center, Indiana University where this work was initiated. 


\newpage
\begin{figure}
\vspace{15cm} \includegraphics{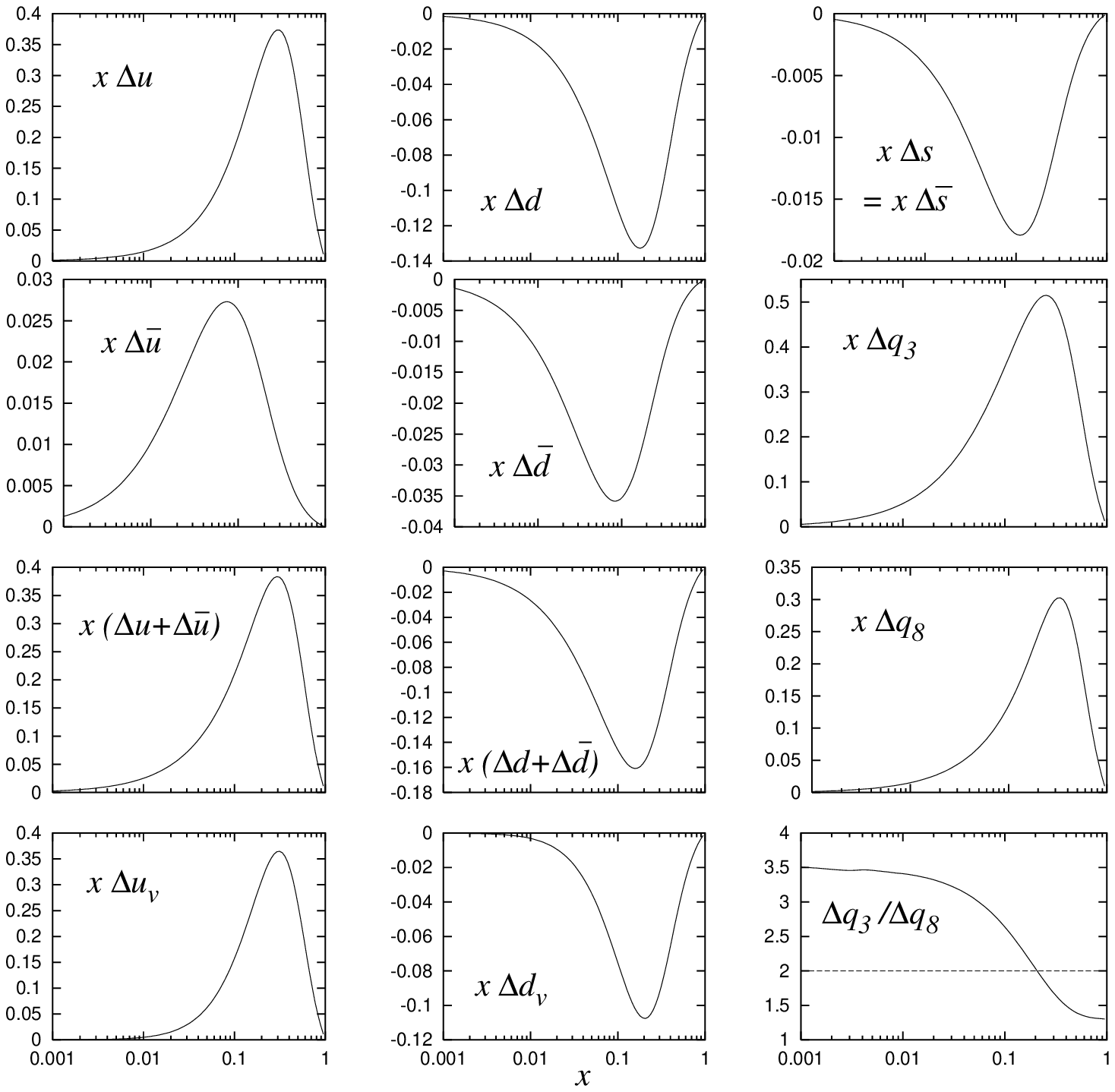}
\caption{Polarized PDFs at $Q^2=Q_0^2~~(=M^2=0.88$ GeV$^2$).}
\end{figure}

\newpage
\begin{figure}
\vspace{15cm} \includegraphics{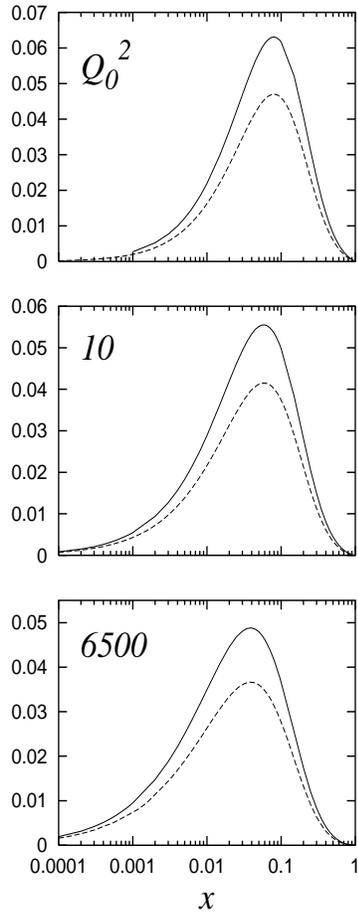}
\caption{Solid curves: $x (\Delta \bar u - \Delta \bar d)$ and
dashed curves: $x (\bar d - \bar u)$. Curves are labelled by
$Q^2$ in GeV$^2$.}
\end{figure}

\newpage
\begin{figure}
\vspace{15cm} \includegraphics{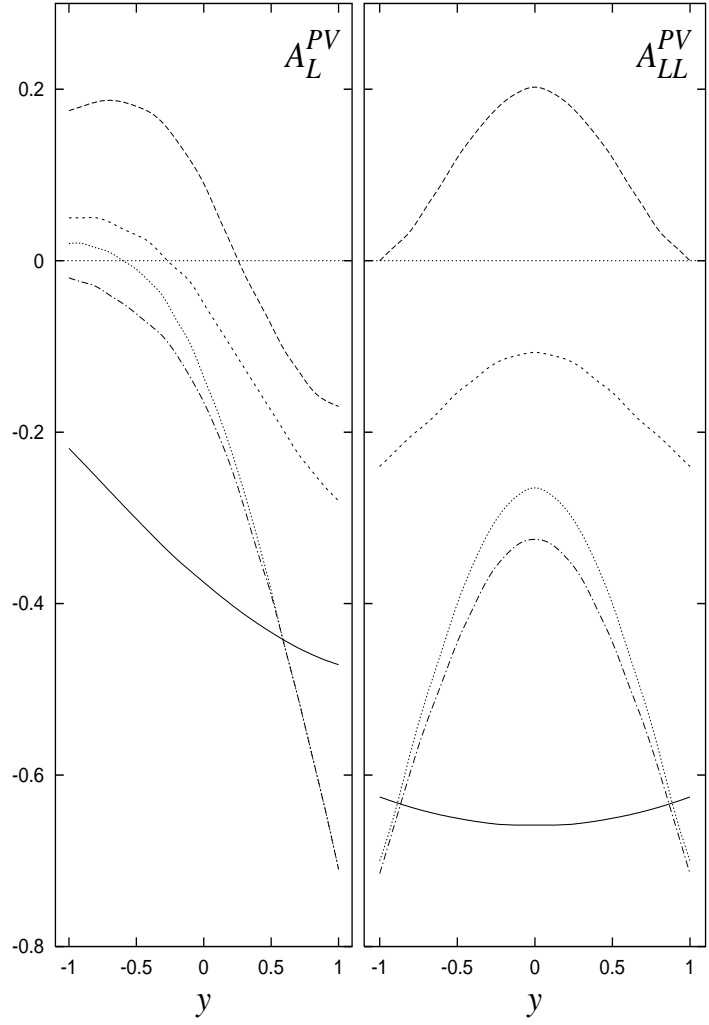}
\caption{Predictions for asymmetries in $W^-$ production in
polarized $pp$ scattering at $\sqrt s = 500$ GeV vs rapidity $y$.
Solid curves: present model. Other curves:
Long-dashed [24],
short-dashed [2],
dotted and dot-dashed [3]
for two separate parameterizations --- as reported in [23].
}
\end{figure}

\end{document}